\documentclass[USenglish,twocolumn]{article}

\ifx\directlua\undefined\ifx\XeTeXcharclass\undefined
  \usepackage[utf8]{inputenc}                           
  \else\RequirePackage[no-math]{fontspec}[2017/03/31]\fi 
  \else\RequirePackage[no-math]{fontspec}[2017/03/31]\fi 
\usepackage[sort&compress,square,numbers]{natbib}
\usepackage[big,online] {dgruyter}
\usepackage{braket}
\usepackage{xcolor}

\usepackage{fancyhdr}

\theoremstyle{dgthm}

\theoremstyle{dgdef}

\begin{document}

\articletype{Research Article}

\author[1]{Ravshanjon Nazarov}
\author[2]{Denis Khanabiev}
\author[3]{Elizaveta Chernysheva}
\author[4]{Alexandra Dudnikova}
\author[5]{Vyacheslav Istomin}
\author[6]{Mikhail Sidorenko}
\author[7]{Jinhui Shi}
\author[8]{Ekaterina Maslova}
\author*[9]{Andrey Bogdanov}
\author*[10]{Zarina Kondratenko (Sadrieva)}
\affil[1]{Qingdao Innovation and Development Center of Harbin Engineering University, Qingdao, 266500, China; and School of Physics and Engineering, ITMO University, St. Petersburg, 197101, Russia}
\affil[2]{School of Physics and Engineering, ITMO University, St. Petersburg, 197101, Russia}
\affil[3]{School of Physics and Engineering, ITMO University, St. Petersburg, 197101, Russia}
\affil[4]{School of Physics and Engineering, ITMO University, St. Petersburg, 197101, Russia}
\affil[5]{Physics and Mathematics Lyceum 239, St. Petersburg, 191144, Russia}
\affil[6]{School of Physics and Engineering, ITMO University, St. Petersburg, 197101, Russia}
\affil[7]{Qingdao Innovation and Development Center of Harbin Engineering University, Qingdao, 266500, China}
\affil[8]{School of Physics and Engineering, ITMO University, St. Petersburg, 197101, Russia}
\affil[9]{Qingdao Innovation and Development Center of Harbin Engineering University, Qingdao, 266500, China; and School of Physics and Engineering, ITMO University, St. Petersburg, 197101, Russia, bogdan.taurus@gmail.com}
\affil[10]{Qingdao Innovation and Development Center of Harbin Engineering University, Qingdao, 266500, China; and School of Physics and Engineering, ITMO University, St. Petersburg, 197101, Russia, z.sadrieva@metalab.ifmo.ru; orcid.org/0000-0001-8299-3226}
\journalname{Nanophotonics}
\journalyear{2025}
\journalvolume{aop}
\title{Bound states in the continuum in a chain of coupled Mie resonators with structural disorder: theory and experiment}
\runningtitle{Running title}

\abstract{We study the impact of structural disorder on a radiative lifetime of symmetry-protected bound state in the continuum (BIC) and a bright mode in a one-dimensional periodic chain of coupled Mie resonators. Through experimental, simulation, and theoretical approach, we reveal an unusual linear decay in the radiative quality factor of the BIC with the increase of the disorder amplitude, contrasting with the quadratic decay  observed in recent studies. We also investigate modes with different symmetries and show that the behavior of the quality factor in a strongly disordered system depends on the mode's multipolar origin. 
Our findings are pivotal for the practical application of BICs, particularly in natural and self-assembled photonic structures where structural disorder plays a crucial role.}


\maketitle

\section{Introduction} 
The concept of bound states in the continuum (BICs) was introduced in quantum mechanics in 1929~\cite{von1993merkwurdige}, and over the last decade, BICs have become a prominent phenomenon in optics~\cite{Koshelev2020Engineering,Xu2019Aug,Vabishchevich2018May,Koshelev2020Jan,Kodigala2017Jan,Ha2018Nov,Huang2020Feb,Yu2021Oct,Hwang2021Jul,Tittl2018Jun,Romano2020Nov,Kravtsov2020Apr,Al-Ani2021Dec}. BICs are
non-radiative states that remain localized in the continuum of propagating waves of the environment. In theory, these states have an infinite radiative quality ($Q$) factor, giving rise to a variety of applications, including nonlinear optics~\cite{Koshelev2020Engineering}, enhancement of nonlinear harmonic generation~\cite{Xu2019Aug,Vabishchevich2018May,Koshelev2020Jan}, optical modulators, lasing~\cite{Kodigala2017Jan,Ha2018Nov,Huang2020Feb,Yu2021Oct,Hwang2021Jul}, sensing technologies~\cite{Tittl2018Jun,Romano2020Nov}, and achieving strong coupling phenomena~\cite{Kravtsov2020Apr,Al-Ani2021Dec}. However, practical aspects such as the finite size of the sample, material absorption, and defects created during the resonator fabrication limit the Q factor of BICs~\cite{Hossein-Zadeh2007Jan,Minkov:13,Biberman:12,Huang2023Jun}. Nanostructure fabrication techniques like ultraviolet~\cite{Mack2018Aug} and electron-beam lithography~\cite{Huang2023Jun} enable precise structural configurations with resolutions down to a few nanometers~\cite{Saifullah2022Sep}; however, lithography often introduces surface roughness. Among various factors, scattering losses due to fabrication imperfections or disorder are the primary limitation on the Q value of BICs, a challenge commonly encountered in the development of high-Q on-chip resonators~\cite{Ishizaki:09,Minkov:13,Biberman:12}.

Structural defects, which can be considered as perturbations introduced to photonic structures, can either deteriorate or enhance the optical properties of a structure~\cite{Madeleine2024}. Recent studies have highlighted that in systems with intentional structural defects, they can improve electromagnetic field enhancement~\cite{Sogrin2023Dec} and localization~\cite{Jo2020Oct}, and stabilize topological states~\cite{Liu2017Nov,Zhou2020Jul,Proctor2020Dec,Roberts2022Dec}. In work~\cite{xiao2018band}, structural disorder leads to the emergence of a bound state in the continuum band in configurations involving multiple chains and layers. Novel
strategies for achieving disorder-resistant ultrahigh $Q$ factors using Brillouin zone folding in BIC metasurfaces have been proposed recently~\cite{Wang2023May}. Moreover,
a merging BIC was found to exhibit significantly reduced radiation losses compared to an isolated BIC within a similarly disordered metasurface~\cite{Jin2019Oct}. Overall,
the development and stability of resonant optical effects resistant to structural defects remain a key challenge in the contemporary nanophotonics research~\cite{nguyen2010total,liu2019disorder,maslova2021bound,Sogrin2023Dec,Zixian2023,kuhne2021fabrication}.

Here, we present both experimental and theoretical analyses of the $Q$ factor for BIC and guided leaky mode in a chain of Mie-resonant ceramic disks with structural disorder.  We show that radiative $Q$ factor decreases linearly with increasing disorder amplitude, in contrast to the one-dimensional periodic structure composed of two layers of dielectric rods considered in our previous work~\cite{maslova2021bound}. We verify our results experimentally in the radiofrequency (RF) range. In contrast to the optical range, RF experiments allow us to neglect the surface roughness of the structural elements and consider only their deviation from the initial position.

%
\section{Regular periodic chain}
\begin{figure}[ht]
\centering
\includegraphics[width=\linewidth]
{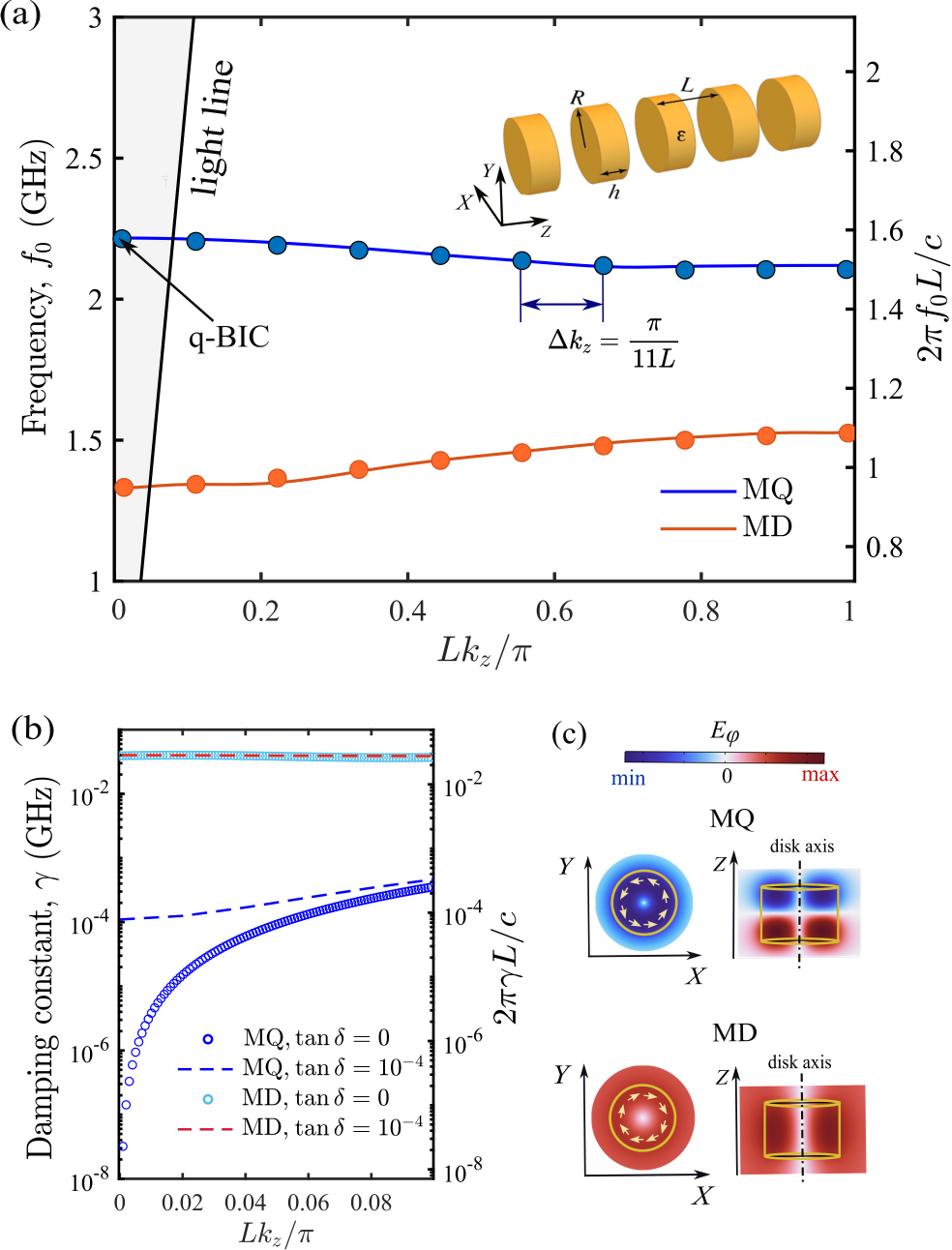}
\caption{(a) Band diagram for the infinite chain of the ceramic disks. The circles show the eigenfrequencies of the finite chain consisting of $10$ disks. The inset shows a 3D model of the chain. (b)  Radiative losses dependencies for eigenmodes of the infinite chain in the presence and absence of loss. (c) Side and front views of the distribution of the azimuthal electric field of the MD and MQ eigenmodes. 
} 
\label{fig1:BandDiagr}
\end{figure}
%
\subsection{Infinite chain}
Let us consider an infinite periodic chain of ceramic disks with period $L$ [see the inset in Fig.~\ref{fig1:BandDiagr}(a)]. According to Bloch's theorem, the electric field of an eigenmode in a coaxial structure can be written as:
\begin{equation}
    \mathbf{E}(r, \phi, z) = \mathbf{U}_{m,k_z}(r, z)e^{\pm i k_zz \pm i m \phi}
    \label{eq1},
\end{equation}
where $k_z$ is the Bloch wave vector, $m$ is the projection of the angular momentum (magnetic quantum number), and $\mathbf{U}_{m,k_z}(r,z)$ is the periodic function of $z$ with period $L$. The modes propagating in the $+z$ and $-z$ directions are degenerate, which leads to $\pm$ signs in the exponential. Similarly, due to the rotational symmetry, modes with $\pm im\phi$ are degenerate. The periodic Bloch amplitude can be expanded into a Fourier series:
\begin{equation}
    \mathbf{U}_{m,k_z}(r, z) = \sum\limits_{n}\mathbf{C}_{m, k_z}^n(r)e^{i 2\pi n z/L},
\end{equation}
where $\mathbf{C}_{m, k_z}^n(r)$ is the Fourier coefficient and $n$ is an integer indicating the diffraction order. In the subwavelength regime, $\lambda>L$, we only have one open diffraction channel, i.e., only the zeroth Fourier coefficient $\mathbf{C}_{m, k_z}^0(r)$ is nonzero and contributes to the far-field.

The modes with $m=0$ have well distinctive polarizations, i.e. they can be divided into purely magnetic (TE) and electric (TM) types, while the modes with $m\neq 0$ have a hybrid polarization~\cite{snyder1983optical}. Ought to this fact the modes with $m=0$ can be symmetry-protected BICs at the $\Gamma$-point, while modes with $m\neq0$ can turn into a BIC only due to fine tuning parameters of the system. An additional advantage of the modes with $m=0$ from the experimental point of view is that they can be selectively excited by the magnetic of electric probe antenna. Therefore, further we will focus only to the modes with $m=0$. We will study two fundamental TE (magnetic) modes representing a chain of coupled magnetic dipoles (MD) and magnetic quadrupoles (MQ). The polarization of the modes is the following $\textbf{E} = (0, E_{\phi}, 0)$ and $\textbf{H} = (H_{r},0,H_{z})$.   

We start our study with numerical simulations of the eigenmodes of the infinite chain composed of ceramic disks with permittivity $\varepsilon = 44$, loss tangent $\tan\delta = 10^{-4}$, disk radius $R = 15$~mm, disk height $h = 20$~mm, and chain period $L = 34$~mm. The fundamental magnetic Mie modes of the disk are magnetic dipole (MD) and magnetic quadrupole (MQ). Their field distribution is shown in Fig.~\ref{fig1:BandDiagr}(c). Due to the coupling of the disks, the MD and MQ bands are formed. Figure \ref{fig1:BandDiagr}(a) shows the band diagram of the infinite chain, where the bottom branch (orange line) corresponds to the MD mode and the top branch (blue line) -- to the MQ mode. The complete band structure accounting other Mie resonances are shown in Fig. S1.  Although the MQ mode belongs to the radiative continuum, in fact, in the $\Gamma$ point, this state is low-radiative with damping constant $\gamma = 10^{-4}$~GHz in the lossy case. However, without material absorption, $\gamma$ tends toward zero in the center of the Brillouin zone [Fig.~\ref{fig1:BandDiagr}(b)]. Taking into account the symmetry of the field distribution, one can conclude that it is a symmetry-protected BIC~\cite{zhen2014topological}. Due to large permittivity, modes are highly localized within the disks; therefore, the total $Q$ factor of BIC is limited by the inverse value of $\tan \delta$, i.e. about 10~000. In contrast, the MD branch exhibits rather large radiative losses [orange line in Fig.~\ref{fig1:BandDiagr}(b)], so it corresponds to the bright mode.

%
\subsection{Finite chain}
In the experiment, we deal with chains of a finite length. In this case, the ends of the chain play the role of partially reflecting mirrors and, thus, the chain can be considered a Fabry-P\'erot resonator. As a result, for a chain of $N$ scatterers, the continuous band of an infinite chain is replaced by a finite set of $N$ Fabry-P\'erot resonances at the frequencies corresponding to the quantized quasi-wave vector $\Delta k_z = \pi/[(N +1)L]$~\cite{Sidorenko2021Mar,sadrieva2018experimental}. Therefore, in a finite chain, a genuine BIC turns into a quasi-BIC with a finite $Q$ factor. The resonant state with the highest Q factor has a quantized wave vector closest to the BIC one. For $N=10$, in Fig.~\ref{fig1:BandDiagr}(a),
eigenfrequencies associated with MD and MQ branches are shown with circle markers, and quasi-BIC appear in the vicinity of the $\Gamma$ point.

%
\section{Effect of disorder}
\begin{figure}
\centering
\includegraphics[width=0.5\linewidth]{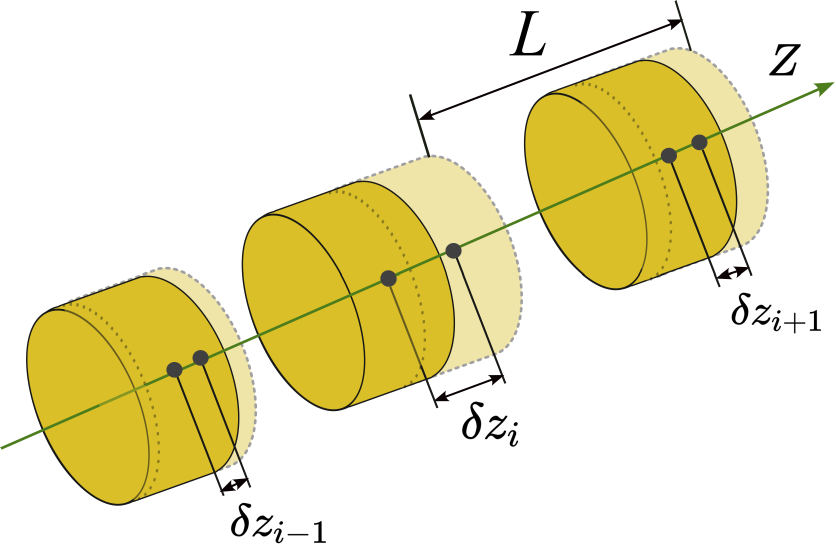}
\caption{Schematic representation of a disordered chain of disks with period $L$. The disorder is induced by shifting the $z$-coordinates of the disks' centers by $\delta z_{i}$, where $i$ is the number of the disk.}
\label{disorder}
\end{figure}

%
\subsection{Numerical analysis}
We investigate the effect of structural disorder in theory and then confirm it with experiments, considering consider finite chains composed of 10, 30 and 50 disks. We introduce uncorrelated disorder to the structure by shifting the coordinates of the disks' centers by $\delta z_i$ along the $z$-axis (Fig.~\ref{disorder}). The shift is given by
\begin{equation}
\delta z_i = \varrho_i L \sigma,
\end{equation}
where $\sigma$ is the disorder amplitude, $\varrho_i$ is a random variable distributed uniformly within the interval $[-1;1]$, and $i$ is the number of the disk. Thus, by applying $\delta z_i$ to the $z$-coordinates of the disks' centers, we provide a random shift and regulate its magnitude by changing $\sigma$.
\begin{figure}
\centering
\includegraphics[width=1\linewidth]{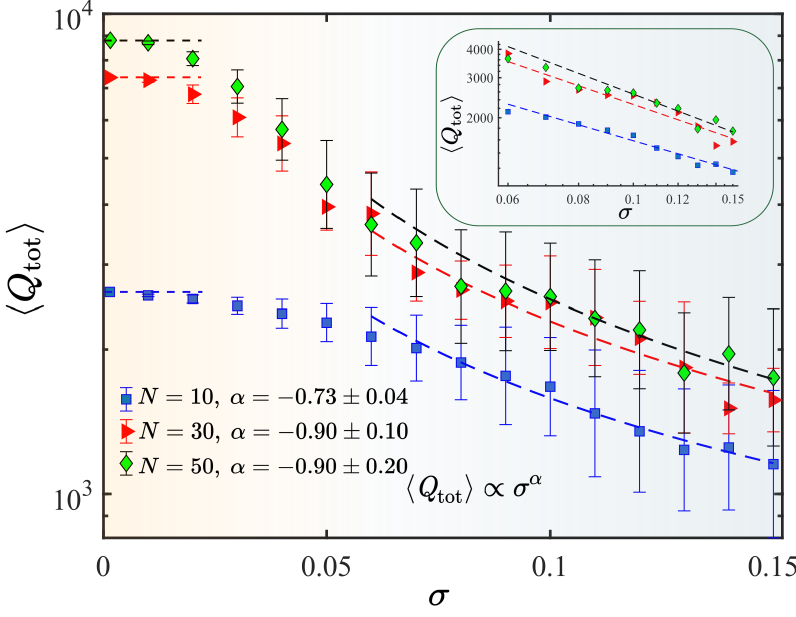}
\caption{ Total $Q$ factor of the quasi-BIC depending on the disorder parameter $\sigma$ for different numbers of disks in the chain $N$. We examined an ensemble of 100 chains for each value of disorder amplitude.The inset shows the $Q$ factor dependecies in doulbe logarythmic scale. 
With the background color, we distinguish the leading channel of losses. For small $\sigma$, the losses due to the finite size dominate; for large $\sigma$, structural disorder plays the main role.}
\label{fig2:Q for var N}
\end{figure}
\begin{figure*}[htb]
    \centering
    \includegraphics[width=\linewidth]{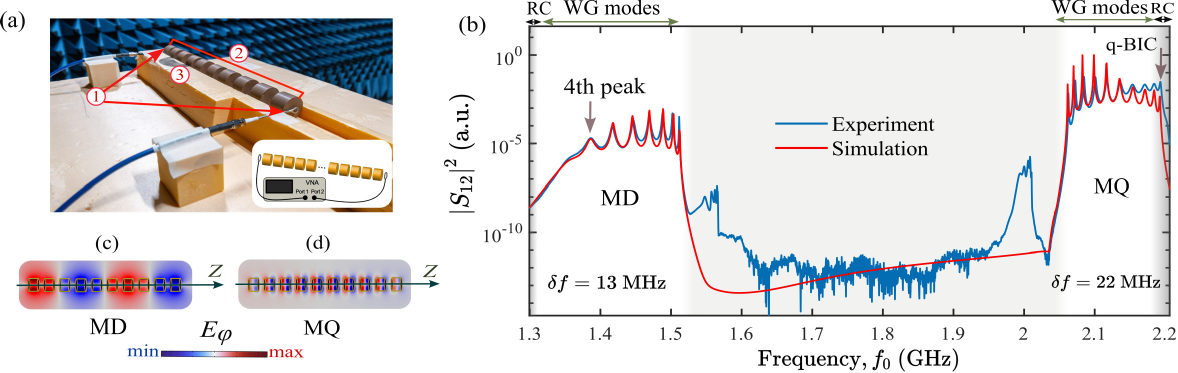}
    \caption{
        (a) Photo and scheme of the setup and samples. 1 - antennas, 2 - ceramic discs, 3 - foam holder. (b) Comparison of simulated (red curves) and experimental (blue curves)transmission spectrum of the 10-disk chain after taking into account the spectral shift $\delta f$. Panels (c) and (d) show the distribution of the electric field in the 10-disk chain for MD and MQ eigenmodes. Grey regions correspond to the radiation continuum, while the region of waveguide (WG) modes is indicated with the background color.}
    \label{fig:setup}
\end{figure*}
To analyze the evolution of the $Q$ factor, we calculate the eigenmodes of the chain. In particular, for a given value of $\sigma$, an ensemble of a hundred randomly generated chains are examined. Among the obtained solutions, we select the fundamental Fabry-P\'erot mode,
i.e. the closest resonance to the $\Gamma$ point, which could be associated with a quasi-BIC. For $N=10$, the fundamental mode is quite easy to find even for a large $\sigma$. As the chain length increases, a more complex behavior is observed: (i) many defect modes with electromagnetic field localized in dimers or trimers see the field distribution in the Section V of Supplementary Materials) appear; (ii) there are two sets of similarly modulated Fabry-P\'erot modes: the first, second, and so on) caused by the field localization at the different ends of the chain. To distinguish proper eigenmodes and perform their statistical analysis, we have developed and implemented a post-processing algorithm with manual control realized in MatLab software. The algorithm employs a 3-sigma rule for a normally distributed random variable, since we observe normally distributed $Q$ factors, see Fig.~S2 in Supplementary Materials. In ambiguous cases, we compared the frequency of a mode with the average values for the previous $\sigma$ using the band diagram. We avoided defect modes with field localized on dimers and trimers, and prioritized those whose field was distributed more uniformly and widely along the length of the chain. Due to the described difficulties, for long chains, we only considered $\sigma$ up to 0.15. 

As shown previously, due to the finite size of a structure, a nonradiative genuine BIC transforms into a quasi-BIC with a finite $Q$ factor~\cite{bulgakov2019high,sadrieva2018experimental}. In addition to radiative losses, we introduce material absorption as a tangent loss of $\tan \delta = 10^{-4}$, estimated from Ref.~\cite{Nenasheva2017}. For high-refractive index ceramics, in which the electromagnetic field is mainly concentrated inside the resonators, the maximum achievable value of $Q$ is limited by the inverse loss tangent, i.e. $10^4$. For instance, in a lossless chain of 50 disks, the $Q$ factor of a quasi-BIC is almost $10^5$, while in the lossy case, it is about $9000$ (see Fig.~\ref{fig2:Q for var N}, green markers). With increasing $N$, the $Q$ factor approaches its limit of $10^4$ defined only by material absorption.

In general, the presence of disorder leads to the mixing of modes with  the different wave vectors~\cite{Poddubny2012,maslova2021bound}.
For an infinitely long periodic chain, weak disorder leads to the slight change of the wave vector, which in turn, 
provides a quadratic decline of radiative lifetime and the $Q$ factor. This is due to the fact that, in the vicinity of the
$\Gamma$ point, the dependence of $Q$ factor from wave vector is quadratic~\cite{zhen2014topological}.
Moreover, a disorder introduced into the system leads to additional radiative losses. Thus, the overall $Q$ factor is determined by
\begin{equation}
    \frac{1}{Q_{\text{tot}}} = \frac{1}{Q_{N}} +\frac{1}{Q_{\text{abs}}} + \frac{1}{Q_{\text{dis}}},
\end{equation}
where losses due to the finite size, material absorption, and parasitic rescattering on structural disorder are summed.  In the case of finite chain, among the observed loss mechanisms, the dominant one can be distinguished. While $\sigma$ is small, the $Q$ factor is approximately constant, and its value is determined by the losses stemming from the finite size and absorption ~\cite{maslova2021bound}. For example, for an ordered chain with $N=10$, $Q$ is 2600, see the blue markers in Fig.~\ref{fig2:Q for var N}.   
 With increasing $\sigma$, the $Q$ factor becomes sensitive to structural disorder.
Starting from $\sigma \approx 0.02$, the $Q$ factor decreases notably. Once the disorder amplitude reaches a threshold of $\sigma \approx 0.05$, its further increase results in a sublinear decay of the $Q$ factor, indicating that the scattering effects due to the structural disorder dominate over those of finite size and absorption. In other words, as the disorder increases, reaching approximately several percent of the period, the intra-band interactions enhance, resulting in mixing of modes and decreasing of the $Q$ factor. In this case, as we show below, the $Q$ factor decay law depends on multipolar origin of the BIC. Notably, the $Q$ factor decay law tends to a linear function  as the number of discs increases.
In contrast to the obtained results, recently it has been shown that in a two-layered one-dimensional periodic structure composed of two arrays of teflon dielectric rods, $Q$ factor decreases quadratically with $\sigma$~\cite{maslova2021bound}. 
Since teflon has a lower permittivity compared to ceramics, we have decided to check whether the permittivity value affects the $Q$ factor decrease law, see Fig.~S4 in the Supplementary Materials. We consider $\varepsilon=15$ and $\varepsilon=6$ and conclude that regardless of its value, $Q$ decays linearly. For smaller values of $\varepsilon$,
the field is localized weakly, and our method cannot define $Q$ factor in the case of disorder properly. It is notable
that in Ref.~\cite{maslova2021bound} the authors consider a q-BIC associated with a magnetic dipole mode (MD),  whereas our study focuses on a magnetic quadrupole q-BIC (MQ). We have calculated the $Q$ factor of a magnetic dipole q-BIC in a ceramic grating with the optical contrast and cross-section similar to those of the chain, see Fig.~S9 in Supplementary Materials. In addition to simulations, we have developed a theoretical model, see Sec.~\ref{Sec:theory}, and obtained the $Q$ factor of the magnetic dipole mode in a chain of lossless circular rods with the same permittivity as for the disks (see Fig.~S10 Supplementary Materials). As a result, we have obtained a near-quadratic decaying of the $Q$ factor of the dipole q-BIC. Thus, within the specified range of permittivity $6\leq\varepsilon \leq 44$, the optical contrast  has no impact on the law governing the decrease in the $Q$ factor. 
Based on the comparison of MD and MQ modes, we can assume
that the $Q$ factor dependence on $\sigma$ is related to the multipolar origin of the q-BIC. It has been shown recently that in a periodic array with a regular asymmetry, the scaling law of the $Q$ factor depends on the origin of the mode~\cite{kutuzova2023quality}. Following this idea, rather strong resistance of the quadrupole q-BIC compared to the dipole q-BIC can be attributed to
the high field localization of the MQ mode, see Fig.~S11 in Supplemental Material.


For large $\sigma$ values, the difference between the average $Q$ factors of chains with different lengths starts to decline, and at $\sigma >0.15$, it becomes negligible, manifesting spatial localization of the mode at a scale less than the length of the structure~\cite{maslova2021bound}.

In addition to uniform distribution discussed above, we have investigated the normally distributed random displacements of the disks center of the lossless chain. In both cases, the $Q$ factor decays almost linearly. More details are provided in Sections II and IV of the Supplementary Materials.
%
\subsection{Experimental study of BIC and non-BIC modes}
\begin{figure}[ht!]
\centering
\includegraphics[width=\linewidth]{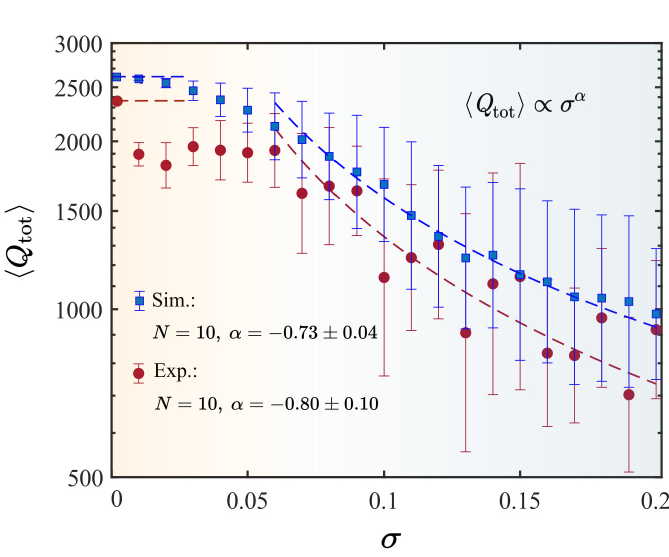}
\caption{Theoretical and experimental dependencies of the total $Q$ factor of a quasi-BIC on the disorder parameter. The markers show the average $Q$, while bars depict the standard deviation. In transmittance measurements, the chain was re-assembled 10 times for each disorder amplitude $\sigma$. In simulations, an ensemble of 100 chains was studied.}
\label{fig3:Exp vs Sim for N=10}
\end{figure}
We considered a chain of 10 cylindrical resonators made of a low-loss microwave ceramic. The nominal permittivity of each resonator is $\varepsilon=44$, and the tangent loss is $\tan \delta=1 \times 10^{-4}$ at a frequency of 1 GHz. Each resonator has a diameter of 30 mm and a height of 20~mm. All the resonators have their axis aligned along a straight line. To control their positions along their common axis, a holder made of a microwave transparent material was designed and fabricated. As the holder material, styrofoam with $(\varepsilon \approx 1.02 - 1.04)$ and negligible losses was chosen. Despite its very low weight, this material is rigid enough has the sufficient rigidity to ensure precise placement of the resonators. A CNC milling machine was used to cut the holder out of a 5 cm thick foam slab.
A cavity was cut out to place the disc resonators coaxially, with varying spacing between them.
\begin{figure} [t!]
\centering
\includegraphics[width=\linewidth]
{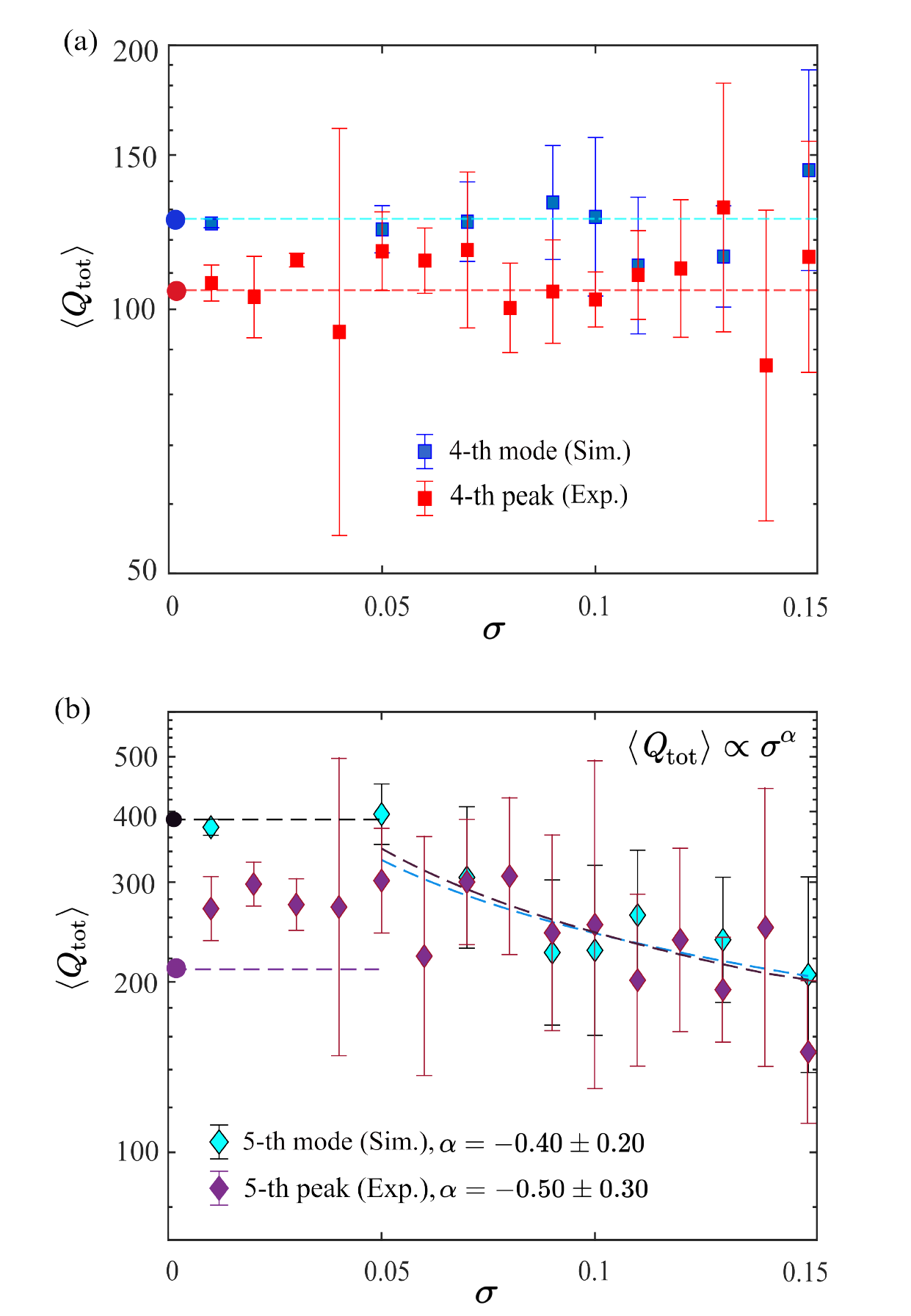}
\caption{ Total $Q$ factor dependencies from disorder parameter for leaky mode. Results on panel (a) corresponds to the 4-th peak, while in panel (b) the fitting results for 5-th peak are listed. In the experiment, 5 different measurements were made for one value of disorder, while we consider 10 different chains during simulation.}
\label{fig5:nonBIC}
\end{figure}

To measure the transmission and reflection coefficients of the structure, a Keysight (Agilent) E8362C Vector Network Analyzer (VNA) was used with a pair of magnetic dipole antennas connected to the VNA ports, see Fig.~\ref{fig:setup} (a). The antennas' axes were aligned with the axis of the disk chain. The size of the magnetic antennas was chosen so that their own resonances (at about 5 GHz) were high above the frequency band of interest (2-3 GHz). Therefore, in the frequency band of 2-3 GHz, their parameters are almost independent of the frequency. Thus, the measured complex values of $S_{21}$ of this setup are proportional to the transmission and reflection coefficients of the resonator chain. 
We needed a large number of frequency samples to ensure sufficient resolution and carefully observe high-$Q$ modes in the transmission spectrum. We decided to take measurements at twenty points in the interval $\sigma \in [0; 0.2]$ with an
increment of 0.01. For each $\sigma$, we made 10 measurements. Then, for each measurement, we plotted the modulus of the vector $S_{12}$ depending on the signal frequency and used these plots to determine the peak in the vicinity of the assumed resonant frequency $f_0$. 
It is important to note that, regardless of the level of disorder, we observe a resonant transmission through the chain. 
Examples of transmittance spectra in the presence of disorder are provided in Fig.~S8 in Supplementary Materials.

A representative measured transmission spectrum for a chain of 10 disks placed equidistantly is plotted in Fig.~\ref{fig:setup}(b). We observe two resonant transmission bands corresponding to the MD and MQ modes.
The resonances lying in the unshaded area correspond to the waveguide modes, and those in the grey area are leaky modes coupled to the radiation continuum. The simulated spectra show good agreement with our experiment, but there are frequency shifts $\delta f = 0.013~\text{GHz}$ and $\delta f = 0.022~\text{GHz}$ for the MD and MQ modes, respectively. Two anomalous peaks within the band gap observed at about 1.55 and 2~GHz in experiment correspond to modes with $m \neq 0$, compare with Fig.~S1 in Supplementary Materials. These modes are excited due to imperfect alignment of antennas. 

First, we focus on the MQ mode turning into BIC. We observe 10 peaks in the spectrum, as it should be for a chain of 10 disks. The last peak, i.e. the highest-frequency one, corresponds to a quasi-BIC, and its field distribution is
plotted in Figs.~\ref{fig:setup}(c,d). The loaded $Q$ factor is extracted as $Q = f_0 / \Delta f$, where $\Delta f$ is the full width of the local maximum in the transmission spectrum at frequency $f_0$. The width and resonant frequency are extracted by fitting the transmission spectrum with the Fano formula~\cite{Limonov2017}. The calculated and extracted $Q$ factors and the
approximation of the obtained dependencies are plotted in Fig.~\ref{fig3:Exp vs Sim for N=10}. As we mentioned above, two regions can be discerned in terms of
the dominant loss mechanism. At low $\sigma$, $Q$ factor is mainly determined by the material absorption and radiative losses due to the finite size of the structure. However, the loaded $Q$ is lower than the calculated one by
$\sim 700$. It can be explained by coupling to antennas and parasitic scattering due 
to non-coaxial arrangement of the disks in the chain~\cite{sadrieva2018experimental}.
As the disorder amplitude increases, the difference becomes less significant, and both theoretical and experimental $Q$ factors decay linearly with respect to $\sigma$.

Next, we consider the MD mode, whose spectrum is shown in Fig.~\ref{fig:setup}(b). Contrary to the MQ mode, here, we see 7 distinct peaks instead of 10, and all of them are in the waveguide mode region, i.e. under the light line. The other three
peaks cannot be determined even in simulation because of their low $Q$ factor. The peaks merge with the background or the $4^{th}$ peak since its shape has notable asymmetry. To analyse robustness of the MD mode against structural disorder, we measure transmittance for a chain with varying amplitude of disorder $\sigma \leq 0.15$. Then we extract the $Q$ factors of the $4^{th}$ and $5^{th}$ peaks and compare them with the predictions from simulation of eigenmodes, see Fig.~\ref{fig5:nonBIC}. In the case $\sigma =0$, the first mode corresponding to the peak with the lowest frequency has a low $Q$ factor, and its value does not change under the disorder. It might indicate a limitation of our numeric calculation. Other modes have a higher $Q$ factor, hence, we can observe the effect of disorder. Specifically, 
the $Q$ factor of the $5^{th}$ peak decays differently from that observed for the quasi-BIC~[Fig.~\ref{fig5:nonBIC}(b)]. 
%

%
%
\section{Analytical model}
\label{Sec:theory}

%
 Here we develop an analytical model that accounts for pairwise interactions between all resonators in the array. This model enables the analysis of the complex eigenfrequencies in a disordered array, as it incorporates the dependence of the coupling constant on the distance between the resonators. 

 Due to the interaction between Mie modes of the disks, resonant transmission occurs. It is clearly seen from Fig.~\ref{fig:setup}(b) that the signal drops dramatically for resonances above the light line. It happens due to the coupling to the radiation continuum. All this can be described in terms of coupled mode theory~\cite{haus1991coupled}.
 
It is well known that the mode amplitude $\psi$ in a system of $N$ resonators can be described by the differential equation as follows~\cite{haus1991coupled,mookherjea2006spectral}:
    \begin{equation}
        \dot{\ket{\psi}}={\hat{M}}{\ket{\psi}},
    \end{equation}
    with matrix elements $\hat{M}$.

\begin{figure} [ht!]
    \centering
\includegraphics[width=0.9\linewidth]{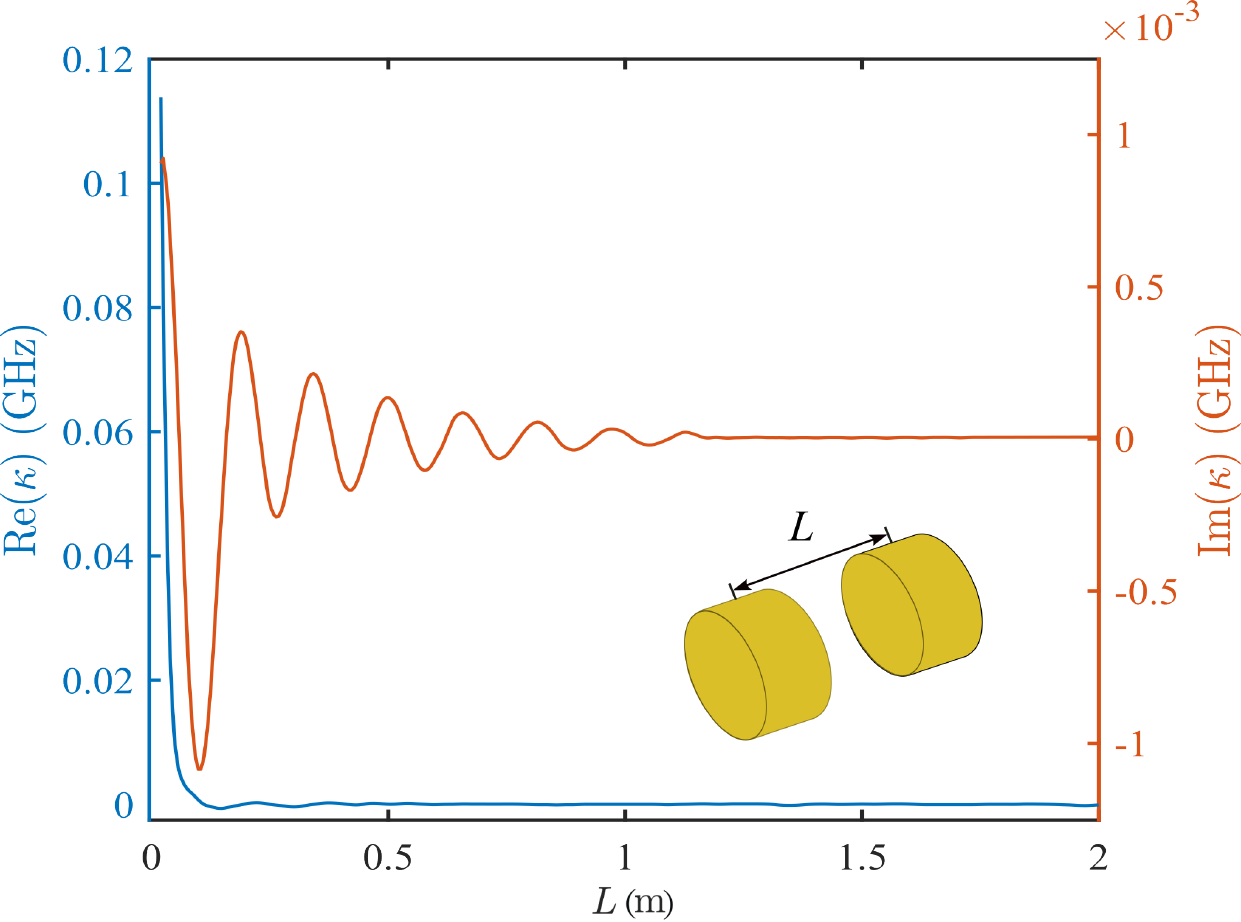}
     \caption{Dependence of coupling coefficient from the distance between two disks. The values are obtained by Eq.~\ref{kappa}}.
    \label{cap}
\end{figure}

After the substitution of $\psi=\psi_0 e^{i\omega t}$, one can get the eigenvalue problem for the matrix $\hat{M}$:
    
    \begin{equation}
        \det(\hat{M}-\omega \hat{I})=0.
    \end{equation}
Matrix  $\hat{M}$ has a form: 

\begin{equation}
        {M_{i,j}}=\Omega \delta_{i,i}+\kappa_{i,j}(1- \delta_{i,j}),
    \end{equation}
where $\Omega$ is the complex resonant frequency of a single resonator, $\kappa_{i,j}$ is the complex coupling coefficient, which depends on the distance between the $i$-th and $j$-th resonators in the chain and $\delta_{i,j}$ is the Kronecker delta. 
Here we assume that the imaginary part of $\Omega$ is only responsible for the coupling to the radiation continuum due to the absence of the material absorption. In our case, the diagonal elements of the matrix $\hat{M}$ are set to the eigenfrequency of the quadrupole mode 
$\Omega=2.141+0.001i$~GHz. Note that we take into account the interaction between all the disks in the chain, i.e. we consider all off-diagonal terms of the coupling matrix. To analyse the coupling coefficients in the case of disorder, we simulated two lossless disks with varying distance $L$ using COMSOL Multiphysics and obtained the frequencies of the symmetric and antisymmetric eigenmodes. Finally, the complex coupling coefficient can be found as follows~\cite{bulgakov2019high}: 
  \begin{equation}
        \label{kappa}{\kappa}=\frac{\Omega_{a}-\Omega_{s}}{2},
    \end{equation}
where the indexes $i,j$ are omitted for convenience. The computed dependence of the complex coupling coefficient on the random distances between the disks $L$ is shown in Fig.~\ref{cap}.

\begin{figure} [ht]
    \centering
   \includegraphics[width=0.9\linewidth]{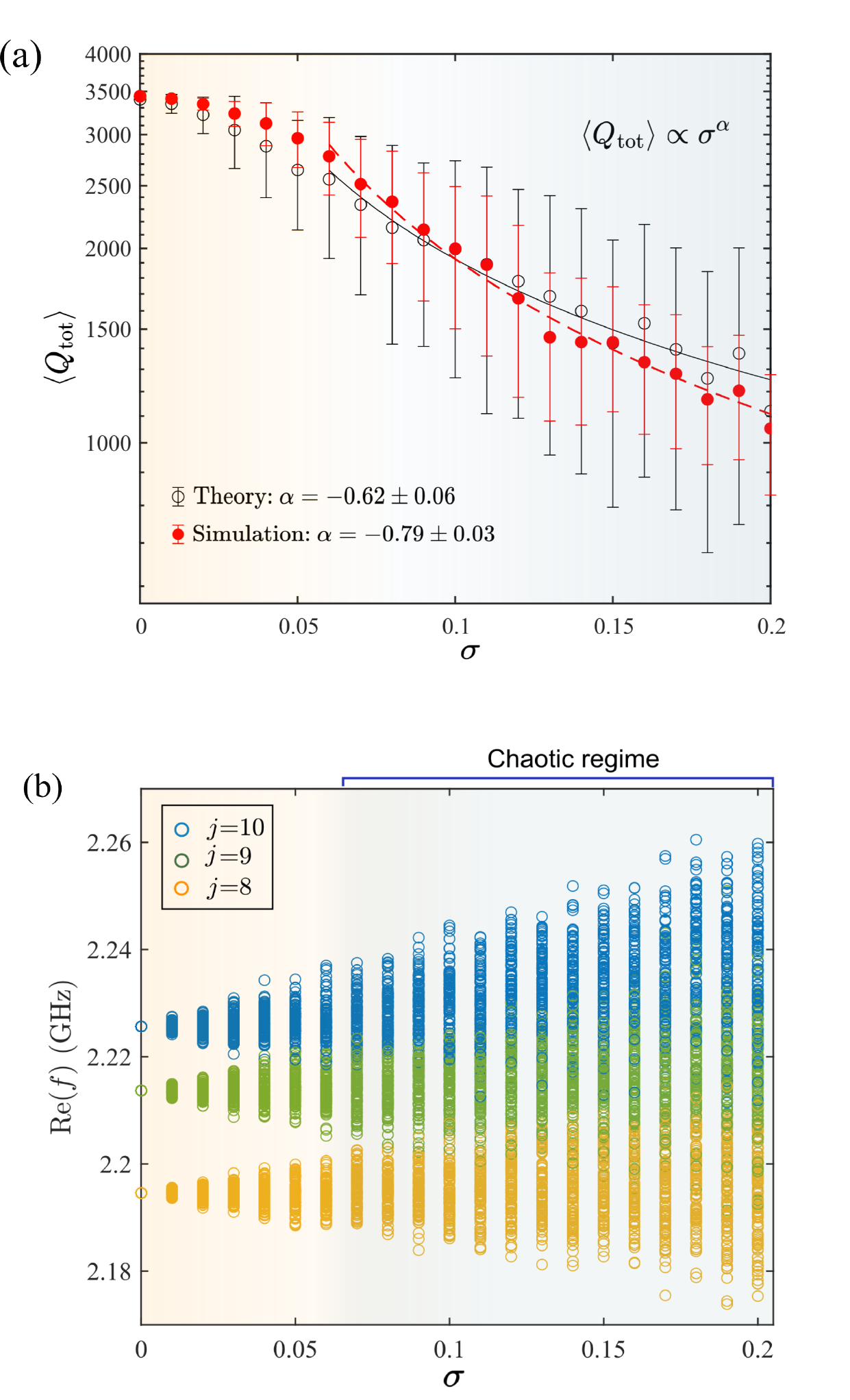}
   
     \caption{ (a) Comparison of $Q$ factor obtained from the solution of eigenvalue problem and simulation for the lossless chain of 10 disks. (b) Behavior of the real part of the 8th, 9th, and 10th eigenfrequencies as a function of the disorder parameter $\sigma$ for 100 realisations.  $j$ denotes the index of the eigenfrequency.The background indicate the dominant loss mechanism: finite-size effects for weak disorder, and structural fluctuation for $\sigma\gtrsim0.5$. In addition, these colors  distinguishes the chaotic regime, where the eigenfrequencies begin to mix, from regular unmixed one.  }
    \label{s}
\end{figure}


The effective Hamiltonian $\hat{M}$ generated for an ensemble of one hundred random samples at the specified amplitude of disorder $\sigma$, is employed to calculate the eigenmode spectrum of the system. The $Q$ factor of the quasi-BIC for the chain of 10 lossless disks, calculated via theoretical model and simulation, is shown in Fig.~\ref{s}(a). The approximation is also presented for comparison. It can be observed that the results are in a good agreement, although the standard deviation is larger than that obtained from COMSOL Multiphysics simulations. This discrepancy can be attributed to the absence of an automatic filtering mechanism for selecting meaningful solutions in the theoretical approach, a feature, that available in COMSOL Multiphysics through mode profile selection. As it will be discussed further, for a larger value of $\sigma$ we observe mixing of 10th and 9th eigenmodes. In order to distinguish the quasi-BIC, we employed either the maximum real part of the frequency filter or the maximum $Q$ factor filter. However, the two filtering algorithms yield the same approximation of $Q(\sigma)$. Finally, both the theoretical approach and the simulation exhibit a consistent sublinear dependence of the $Q$ factor on the disorder amplitude. 

Furthermore, we examine the behaviour of the fundamental mode and its two neighbouring modes in relation to the disorder parameter. As illustrated in 
Fig.~\ref{s}(b), for values of $\sigma$ below approximately 0.05, these modes remain distinct. However, as the disorder increases, a chaotic regime emerges, and the fundamental mode ($j=10$) predominantly mixes with the next Fabry-P\'erot mode ($j=9$). This provides an explanation for the substantial decrease of the $Q$ factor, for $\sigma$ exceeding the threshold value, as demonstrated in both the theoretical and experimental results, see Fig.~\ref{fig3:Exp vs Sim for N=10}. 
\section{Anderson localization}

  Anderson localization, characterized by the emergence of localized states, is a remarkable phenomenon observed in disordered systems~\cite{anderson1958absence,poddubny2012fano,wiersma1997localization}. It has been well studied in various physical systems, especially in 1D photonic structures~\cite{fernandez2014beyond,tarquini2017critical,lahini2008anderson}. In general, disorder disrupts the effective transport of energy through the system, leading to suppressed transmission efficiency~\cite{alu2010effect}. In the case of finite structure, the localization effect occurs when the field becomes confined to a region smaller than the system's characteristic size~\cite{segev2013anderson}. This results in the formation of localized states for which the transmitted field decays along the chain~\cite{poddubny2012fano}.
  
\begin{figure} [t!]
    \centering
   \includegraphics[width=0.9\linewidth]
   {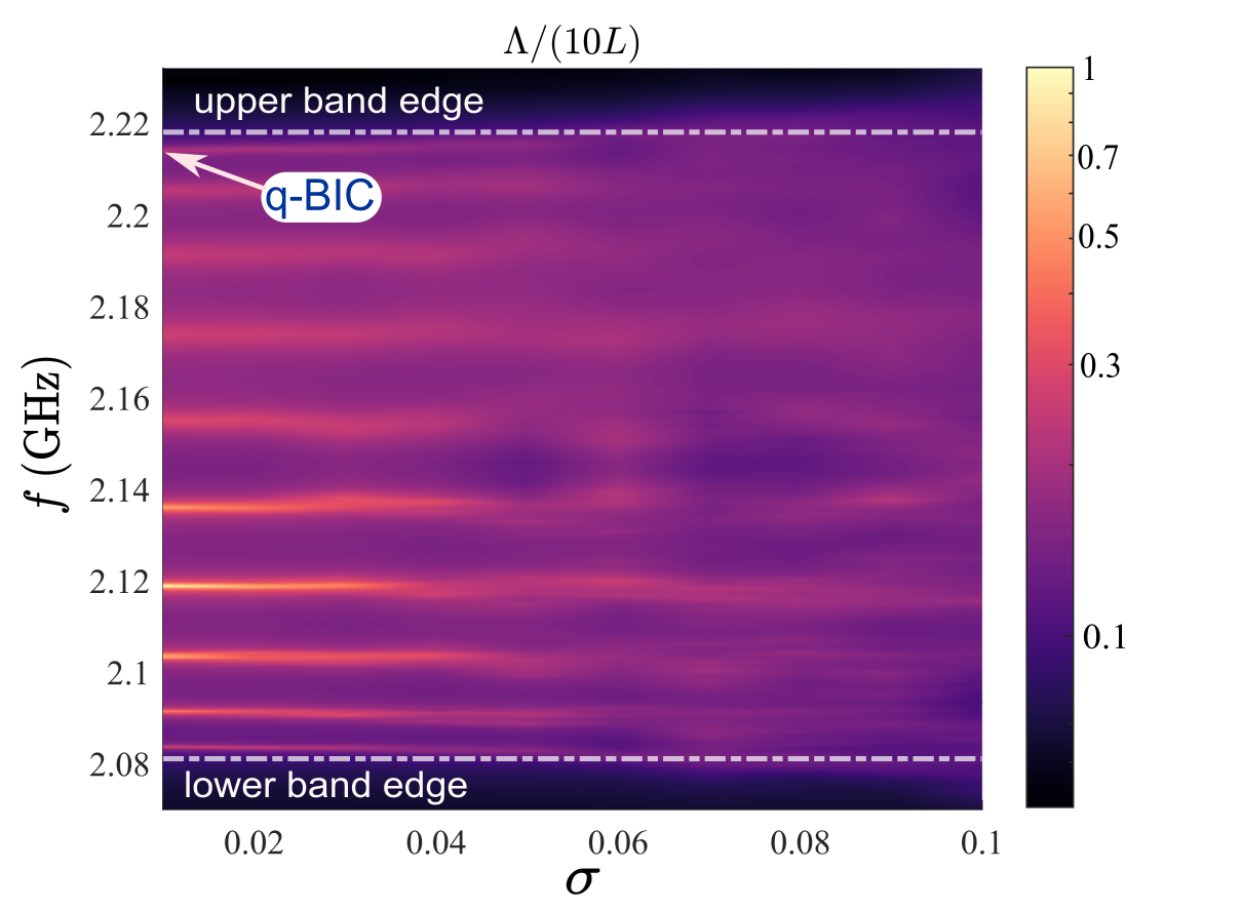}
     \caption{ Numerically calculated localization length for the chain of 10 disks. The map highlights the upper and lower band edges and indicates the frequency of the q-BIC. }
    \label{loclength}
\end{figure}

Here we examine the influence of disorder on the transmitted field and the emergence of localized states in a disordered chain. For this,
we calculated the evolution of a \textit{localization length} $\Lambda$ determined by transmittance $T$ as follows~\cite{mcgurn1993anderson}:

  \begin{equation}
     \frac{1}{\Lambda}=- \frac{{\langle\ln T \rangle }}{NL},
    \label{eq14}
\end{equation}
where $N$ in our case is equal to 10.
Figure~\ref{loclength} shows the localization length as a function of $\sigma$. It is possible to distinguish 10 bright lines, which correspond to the 10 Fabry-P\'erot modes. Among them, quasi-BIC is observed at the highest frequency. 
 Moreover, the behavior of  localization length indicates  the presence of two regimes [Fig.~\ref{loclength}]. Namely, at small $\sigma$ the resonant modes are well separated, while as the disorder increases, they start to mix. At the first regime, the localization length is comparable to the chain length or longer. The localization length is physically limited by both the finite chain size and the presence of non-zero material absorption. At the second regime, the localization length decreases dramatically, indicating the emergence of Anderson-like localization~\cite{mcgurn1993anderson,wiersma1997localization,Poddubny2012}. This, in turn, leads to the appearance of the spatial localized field along the disordered chain~\cite{john1991localization}. This behavior of the modes is the same as shown in Fig.~\ref{s}(b).

Furthermore, the localization length assumes small but non-zero in the vicinity of the band edge within the band gap [Fig.~\ref{loclength}]. The band gap facilitates the formation of a highly localized field in a small region of the disordered system~\cite{mcgurn1993anderson}. However, it is important to note that, irrespective of the level of disorder, both the simulation and the experiment demonstrate a resonant transmission through the chain, as illustrated in Fig.~S13 in the Supplementary materials. This fact serves to prove that our investigation is focused on the destroying of BIC rather than the spatial localization of the field in disordered media.

\section{Conclusion}
In conclusion, we have studied how uncorrelated structural disorder affects a symmetry-protected at-$\Gamma$ BIC and a non-BIC mode in a one-dimensional periodic array composed of ceramic disks. In the experiment, we have selectively excited
magnetic octupole and magnetic dipole modes with zero orbital angular momentum and measured the transmission spectra using coaxially placed loop antennas. We have extracted the $Q$ factor from the experimental data for arrays with 10 disks, revealing different asymptotic depends of the $Q$ factor on the disorder amplitude for the symmetry-protected BIC and non-BIC modes, respectively. Moreover, coupled mode theory predicts a sublinear decay of the $Q$ factor, which is corroborated by both our numerical simulations and experimental results. The behaviour of the $Q$ factor for the magnetic dipole mode has been examined, and a well-known quadratic decay has been revealed.  Therefore, we can conclude that, in the case of strongly disordered finite chain, where neighboring modes start to mix and Anderson localization appears, the intra-band interaction became significant, and the $Q$ factor decreasing depends on the multipole origin of the quasi-BIC.  Additionally, theoretical analysis predicts the range of disorder amplitude at which the chaotic regime emerges, which aligns with the estimated value from the simulations. 

\begin{acknowledgement}
The authors acknowledge useful discussions with Professor Yuri Kivshar, Professor Alexander Poddubny and Professor Mikhail Rybin.
\end{acknowledgement}

\begin{funding}
The work was supported by the Priority 2030 Federal Academic Leadership Program and the Russian Science Foundation (Grant 23-72-10059). 
\end{funding}

\begin{authorcontributions}
All authors have accepted responsibility for the entire content of this manuscript and approved its submission.

\end{authorcontributions}

\begin{conflictofinterest}
Authors state no conflict of interest.
\end{conflictofinterest}

\bibliographystyle{ieeetr} 
\setcitestyle{numbers}
\bibliography{bibliography}

\end{document}


\preprint{APS/123-QED}
\title{Supplemental Material\\
for\\
``Bound states in the continuum in a chain of coupled ceramic disks with structural disorder: theory and experiment ''}
\maketitle
\onecolumngrid
\section{Band diagrams}
\begin{figure}[ht]
    \centering
    \includegraphics[width=0.75\linewidth]{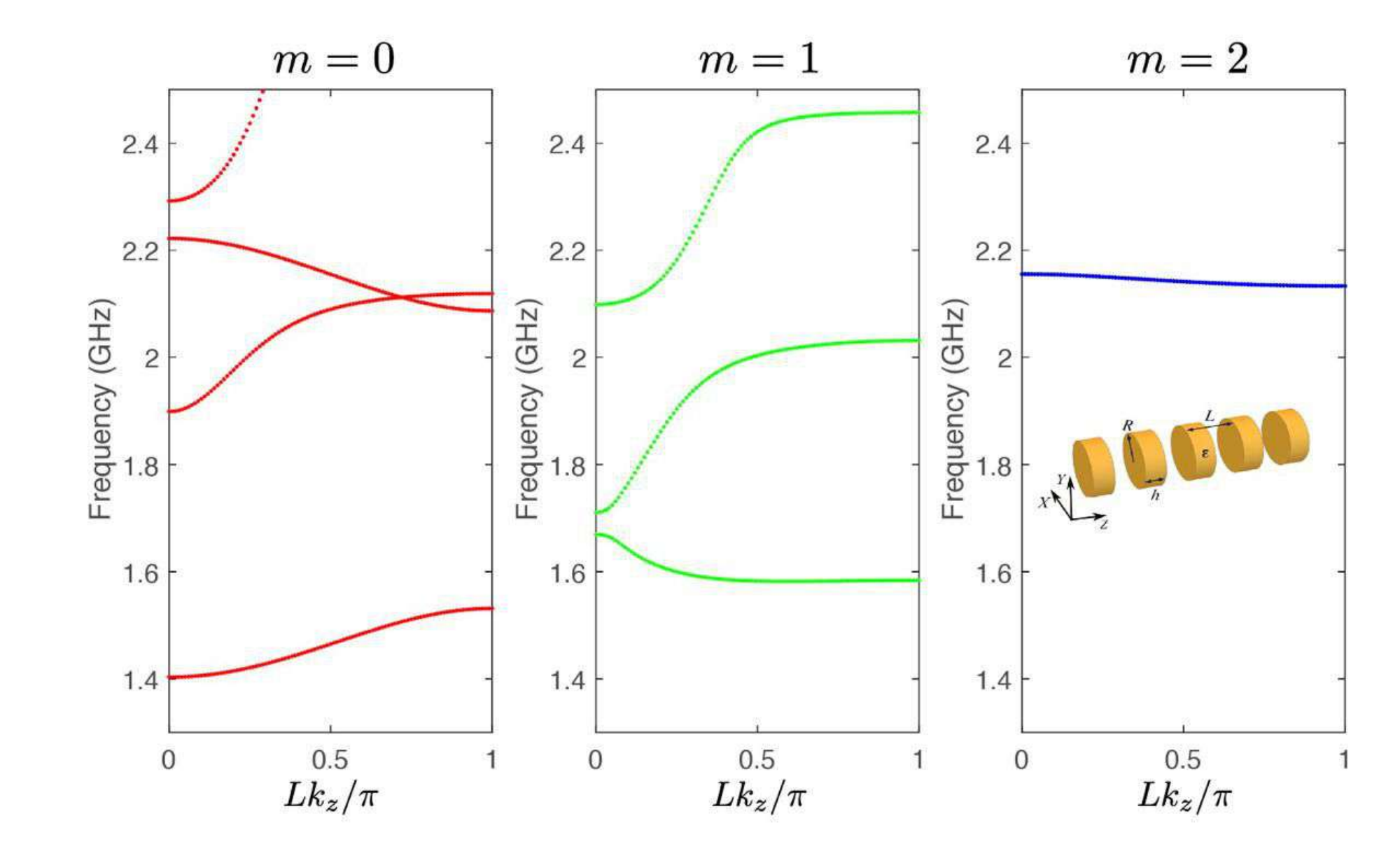}
    \caption{Band diagrams for $m=0,1,2$ in the frequency range of interest.}
    \label{figS1:bands}
\end{figure}

\section{Distribution of quality factor}
\begin{figure}[ht]
    \centering
    \includegraphics[width=\linewidth]{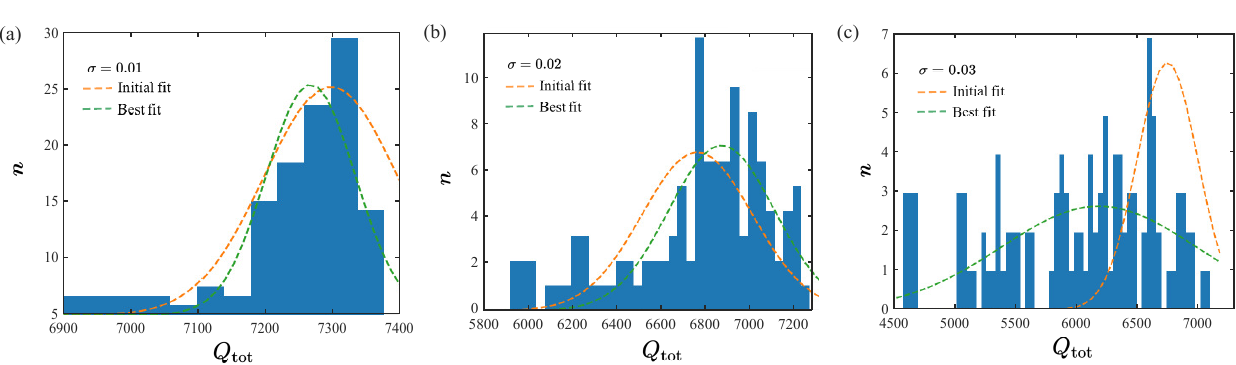}
    \caption{Distribution of the total $Q$ factor and its approximation by the normally distributed function. The chain is composed of 30 disks.}
    \label{figS1}
\end{figure}
\begin{figure}[ht]
    \centering
    \includegraphics[width=0.45\linewidth]{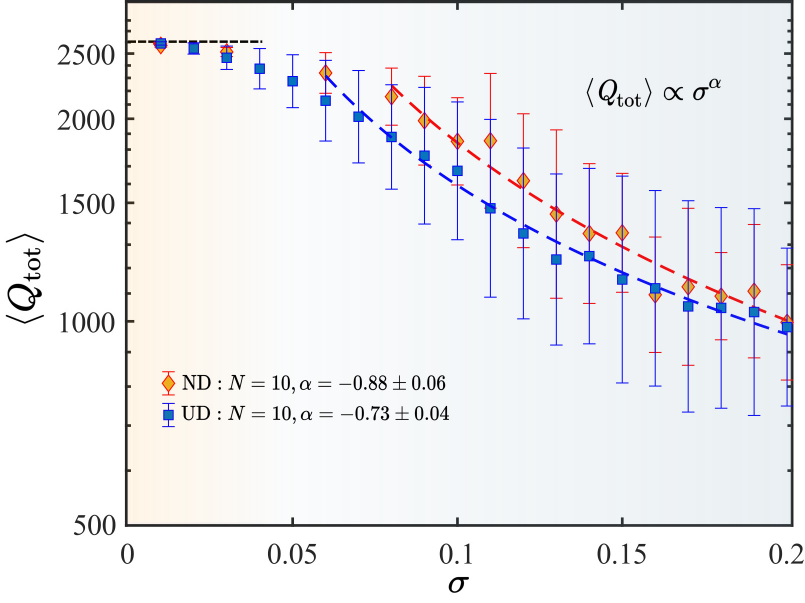}
    \caption{Dependencies of the average value of total $Q$ factor from disorder parameter for normally distributed (ND) and unitary distributed (UD) random coordinates of the disk's center.}
    \label{figS1.1}
\end{figure}

The distribution of the obtained total $Q_{\text{tot}}$  factor does not change significantly with the value of disorder parameter $\sigma$. Indeed, as it is shown in Fig.~\ref{figS1}, with $\sigma$ increasing, $Q_{\text{tot}}$ preserves a normal distribution. For the larger values of $\sigma$, $Q_{\text{tot}}$ may deviate from the normal distribution. 

Next, we checked whether the distribution of a random shift of disks' centers affects the average value of $Q$ factor. We compared normally and unitary distributed random shifts. For the small values of $\sigma$, there is no difference between normally (ND) and unitary distributed (UD) coordinates (Fig.~\ref{figS1.1}). In contrast, with the $\sigma$ increasing, obtained values of $Q$ become slightly different. However, this difference do not change much the decay law. It should be noted that we studied the ensemble of 10 chains, so some discrepancy can appear due to the small amount of chains.
For the larger values of $\sigma$, the behavior of the average of $Q_{\text{tot}}$ in both case is the same.

\section{Various values of dielectric permittivity}
\begin{figure}[ht!]
    \centering
    \includegraphics[width=0.75\linewidth]{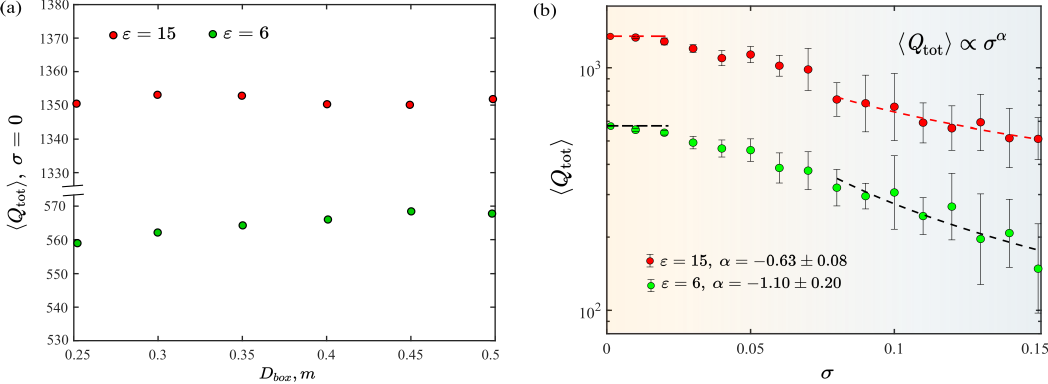}
    \caption{(a) Total $Q$ factor versus the box size $D_{box}$ for different values of disks permittivity $\varepsilon$ in the absence of disorder. (b) Dependencies of the average value of total $Q$ factor from the disorder parameter for $\varepsilon=15$ and $\varepsilon=6$. During the simulation ensemble of 10 chains were studied.}
    \label{figS2}
\end{figure}

In order to check whether the decrease of the dielectric permittivity $\varepsilon$ of disks affects the decay law $Q \sim \sigma^{-1}$, we provide the similar series of simulations for lower values of $\varepsilon$ then in the main text. With COMSOL Multiphysics and MatLab software, we analyse ensembles of 10 chains per each value of amplitude $\sigma$. The chain is composed of 10 disks. Figure~\ref{figS2}(b) shows average $Q$ factor for $\varepsilon$ equals to 6 and 15. The lowest value of $\varepsilon$ was chosen based on the assumption that eigenmodes should be spatially localised and do not interact with the perfectly matched (PML) layers surrounding simulation area even in the presence of disorder, see Fig.~\ref{figS2}(a). The slope of both lines is close to 1, similarly to that one for $\varepsilon=44$ considered in the main text. Thereby, regardless the dielectric permittivity of the disks, structural disorder results in linear decaying of the $Q$ factor with $\sigma$.

\section{Influence of the material losses on quality factor in the disordered chain}

\begin{figure}[ht!]
    \centering
\includegraphics[width=0.45\linewidth]{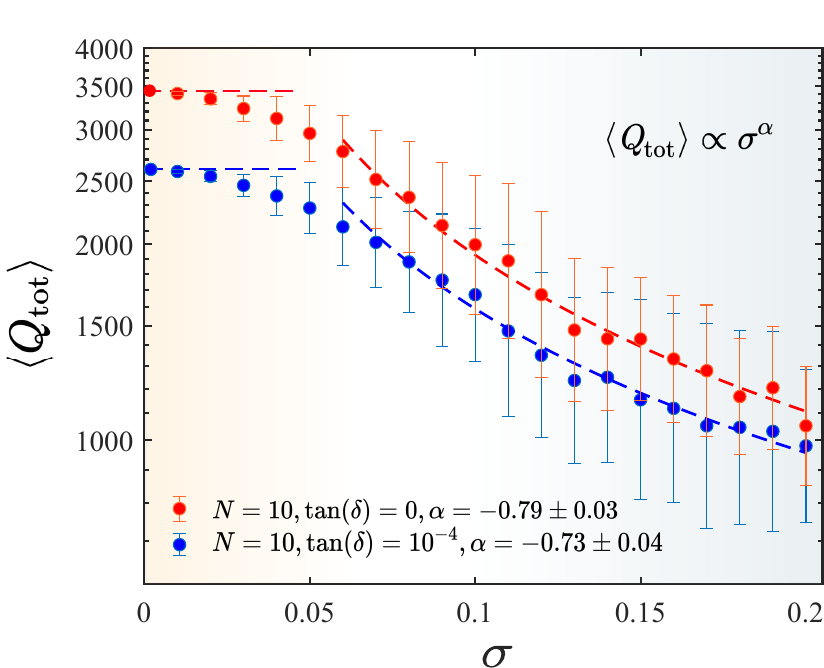}
    \caption{Comparison of average $Q$ factor dependencies from the disorder parameter at the 10-disk chain in case of presence and absence of material losses.}
    \label{figS8}
\end{figure}

In order to observe the effect of the material losses on the $Q$ factor behavior, we studied modes of chain of 10 disks. For a particular value of $\sigma$, we examined 100 randomly generated chains. The obtained results are depicted in Fig.~\ref{figS8}. Obviously, at any value of $\sigma$, $Q$ factor is larger for the lossless chains then for the lossy ones. However, in the lossless case, $Q$ factor decreases following the similar decay law as in the lossy case. 

\section{Electromagnetic field distribution in the chain with disorder}

\begin{figure}[h]
    \centering
\includegraphics[width=0.6\linewidth]{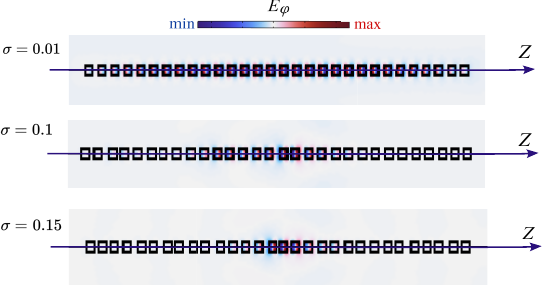}
    \caption{Electric field distribution in the 30-disk chain for the different value of disorder $\sigma$. While the top panel shows the fundamental Fabry-Perot mode, the bottom ones show the defect modes.}
    \label{figS3}
\end{figure}
\begin{figure}[ht]
    \centering
\includegraphics[width=0.45\linewidth]{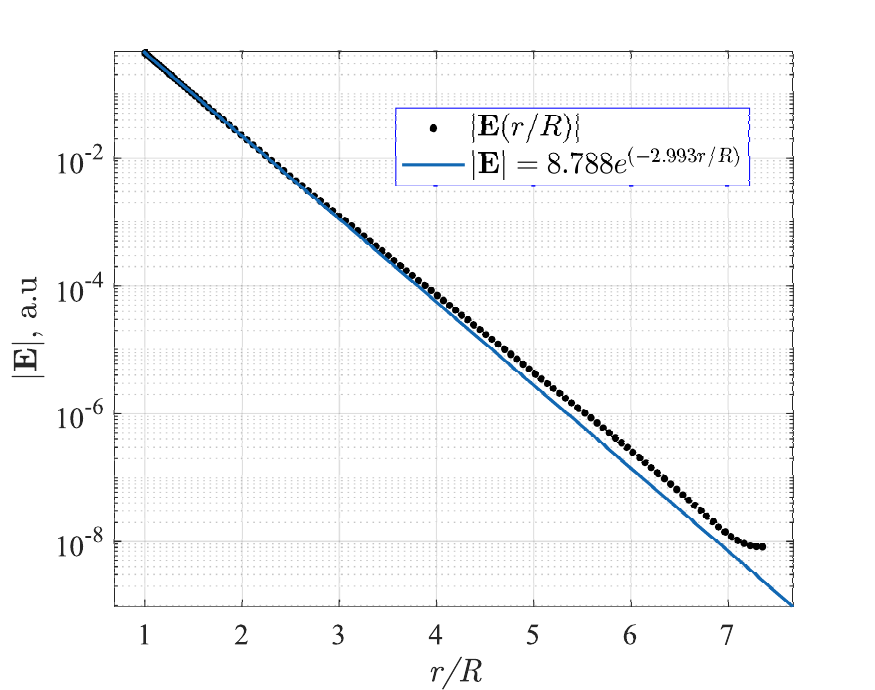}
    \caption{Normalized BIC's field dependence from the distance outside the disk in the case of infinite periodic chain.}
    \label{figS4}
\end{figure}

The increasing of the chain's length leads to the more complex behavior of the mode. Figure \ref{figS3} shows the appearance of localised defect modes in dimers or trimers in the 30-disk chain. Moreover, as amplitude of the disorder increase, these localised modes became narrowly distributed along the length of the chain. During simulations, we exclude the defect modes.
Outside the disk, the field of a quasi-BIC decays exponentially with the distance. This behavior of the mode is shown in Fig.\ref{figS4}. The simulation results (black dots) indicate that for the distance $r/R>4$, the field distribution deviate from the approximation curve. Moreover, when $r/R\geq7$, electric field became constant. This behavior of the field is related to the limits of the accuracy of simulation method.
%
\section{Transmission spectra of disordered chains}
\begin{figure}[ht]
    \centering
\includegraphics[width=0.45\linewidth]{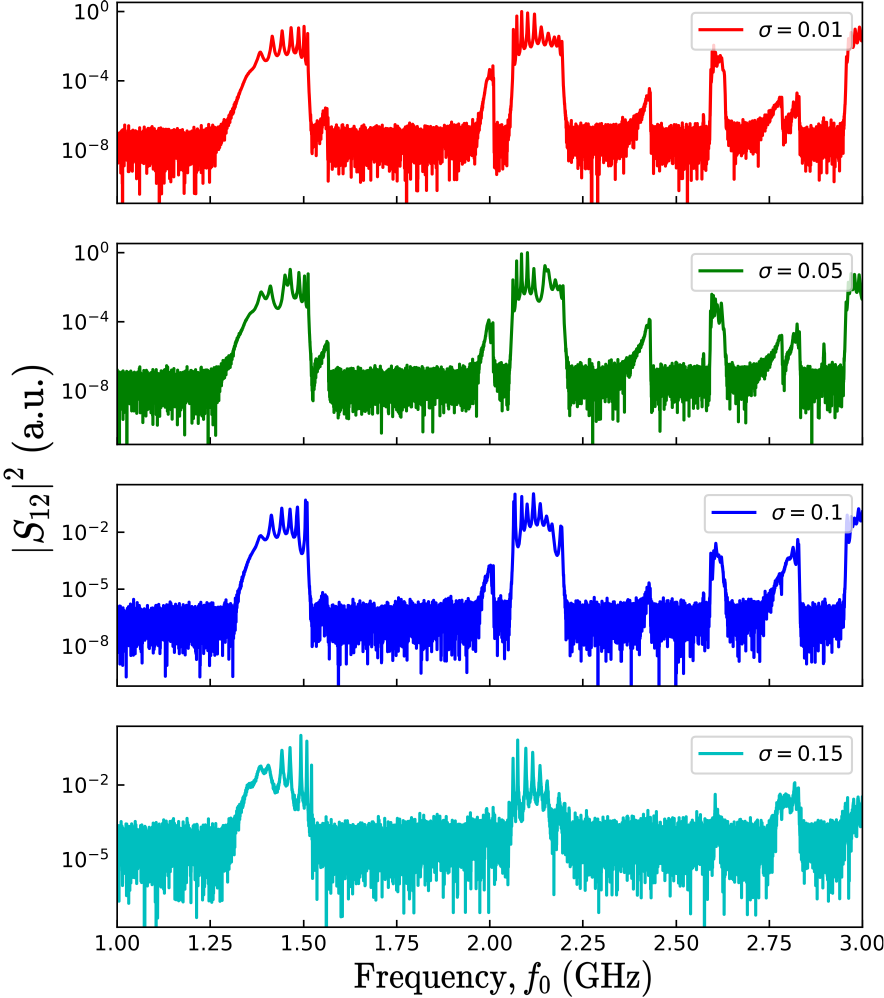}
    \caption{Measured transmission spectrum of the 10-disk chain with different value of disorder parameter $\sigma$. }
    \label{figS5}
\end{figure}

Undoubtedly, the presence of the disorder affects the spectral response of the system. It can be clearly seen in Fig.\ref{figS5}, which illustrates the evolution of the transmittance spectrum with the disorder amplitude. With the $\sigma$ increasing, the amplitude of the response decreases, while level of noise increases. The bands appeared at about 3~GHz vanishes. However, we observe the modes of interest at the same frequencies regardless the value of $\sigma$.   

\section{Multipolar origin of $Q$ factor dependencies from the disorder}
\begin{figure}[ht]
    \centering
\includegraphics[width=0.45\linewidth]{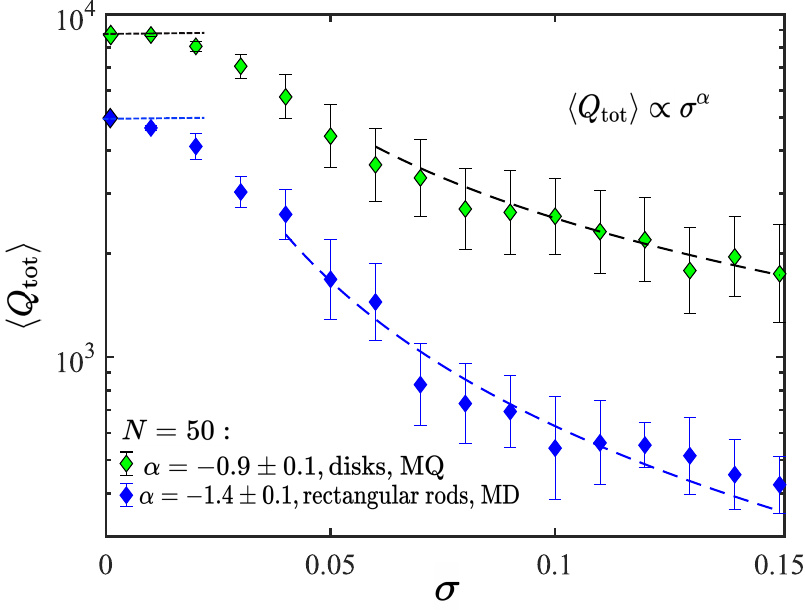}
    \caption{Dependencies of the average value of total $Q$ factor from disorder parameter for quadrupole (MQ) and dipole (MD) q-BICs. The quadrupole q-BIC is observed for the chain of disks, while the dipole q-BIC appear in the finite array of the rods with rectangular cross-section. The number of disks and rods is equal to 50.}
    \label{figS6}
    
\end{figure}

In the previous research, it was shown that the $Q$ factor decays quadratically with the disorder parameter [S1]. The authors studied q-BICs for the dipole modes of the double array of dielectric rods. Since here we consider a quadrupole q-BIC and obtain $Q \sim \sigma^{-1}$, 
we suppose that the $Q$ factor dependence from $\sigma$ is related to the multipolar origin of q-BICs. In order to 
test this hypothesis, we calculate $Q$ for the dipole q-BICs in the finite array of infinitely long ceramic rectangular rods. The height of the rods is equal to the disk's diameter, while the width is similar to the disk's height. Here, we have chosen a grating, since dipole modes of the chain considered in the main text does not support BICs due to the symmetry requirements, see~Fig.1(c). The finite array consists of 50 rods. For a given value of $\sigma$, we calculated 10 randomly generated finite arrays. Figure~\ref{figS6} shows the behavior of total $Q$ factor from $\sigma$ for two q-BICs. One can see that the $Q$ of the dipole (MD) q-BIC decays faster then the quadrupole q-BICs (MQ). We suppose that some discrepancy between expected quadratic decay law and obtained $Q \sim \sigma^{-1.5}$ can appear due to the small amount of chains. 

\begin{figure}[ht]
    \centering
\includegraphics[width=0.8\linewidth]{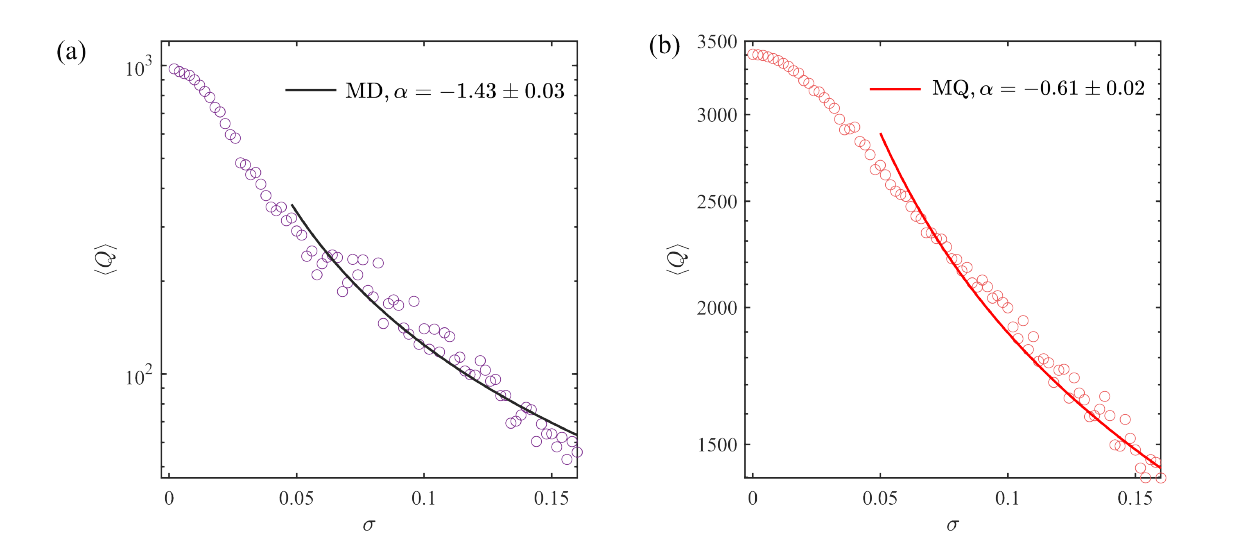}
    \caption{Theoretically obtained dependencies of the mean value of the total $Q$ factor from the disorder parameter (a) for a dipole mode (MD) in a finite chain of 10-circular rods and (b) for a quadrupole mode (MQ) in a chain of lossless 10 discs. The values averaged over 300 realizations.}
    \label{figS62}
\end{figure}

In order to verify the obtained results, we apply the Coupled Mode Theory (CMT) to both dipole and quadrupole modes. A detailed description of the applied theory is provided in the main text. In this analysis, we consider a lossless finite array of 10 circular rods with the same permittivity as that of the disks. The period of the array is 600 nm, while the radius is equal to 225 nm. In the case of an infinite periodic array, we found that a magnetic dipole mode exhibited a at-$\Gamma$ BIC at a frequency of 121.4 THz. As it shown at Fig.~\ref{figS62}, the decay of the $Q$ factor is similar to that observed value in the simulation results. Consequently, it can be concluded that the decay of the $Q$ factor is independent of the geometry and can be explained in terms of multipoles.

\begin{figure}[ht!]
    \centering
\includegraphics[width=0.7\linewidth]{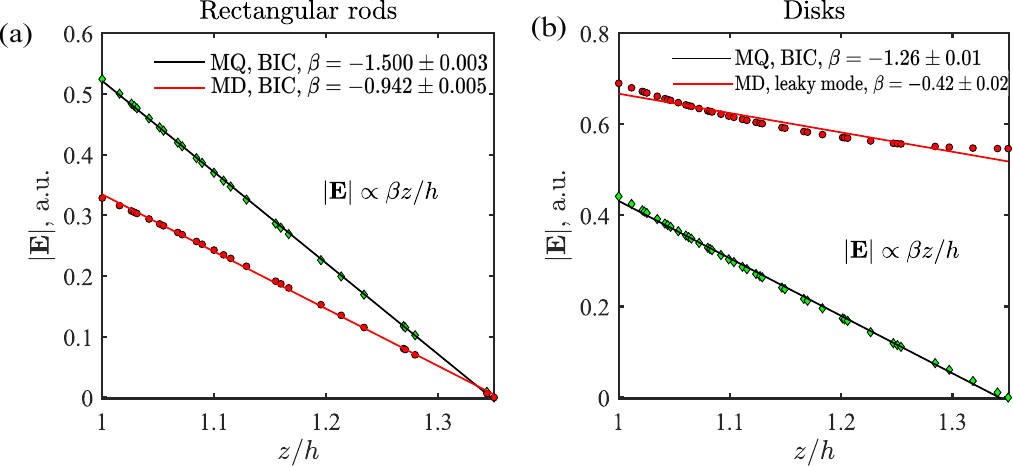}
    \caption{Evolution of the field along the periodic direction $z$ for quadrupole (MQ) and dipole (MD) modes.}
    \label{figS7}
\end{figure}

Obviously, $Q$ factor depends on the localisation of the field in the resonator. 
Figure ~\ref{figS7} shows the field behaviour along the direction of periodicity $z$ within a unit cell. We examine magnetic quadrupole (MQ) and dipole (MD) modes of the grating and chain of disks. One can see that the field associated with BIC decays linearly, while the leaky MD mode of the chain shows nonlinear dependence. Moreover, BICs associated with MQ modes decays faster manifesting about strong localization compared to MD-BIC in the grating.

\section{Dependencies of total $Q$ factor decay law from the chain length}

\begin{figure}[ht]
    \centering
\includegraphics[width=0.9\linewidth]{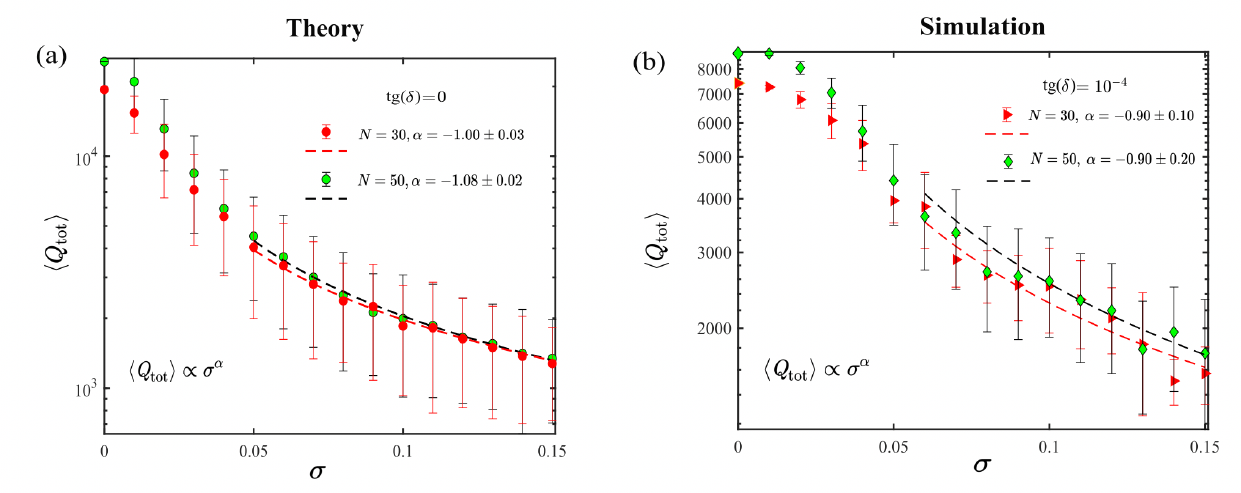}
    \caption{(a) Theoretically obtained dependencies of the mean value of the total $Q$ factor from the disorder parameter for a 300 different realizations; (b) Simulation results for a 100 realizations.}
    \label{figS12}
\end{figure}

 In the main text from the simulation results, we conclude that with the increment of  the chain length, the total $Q$ factor dependencies from the disorder are close to $Q_{tot}\sim\sigma^{-1}$ . In order to verify this, we theoretically calculate the $Q_{tot}$ for $N=30$ and $N=50$ lossless disks. As it is shown at ~\ref{figS12}, the coefficient of total $Q$ factor  decay from the disorder amplitude is close to the  $\alpha=1$. 
\section{Localisation length}
Figure 9 from the main text demonstrates that the mixing of eigenfrequencies is facilitated by the emergence of localized states. A key parameter characterizing these localized states is the localization length. Physically, the localization length can be interpreted as the decay length of the transmittance outside the chain, therefore it can be found as:
  \begin{equation}
    \frac{1}{\Lambda}=-\lim_{N\to\infty} \frac{\ln{\langle T \rangle }}{NL},
    \label{eq14}
\end{equation}
where $\Lambda$ is the localisation length, $NL$ is the length of the chain and $T$ is transmittance [S2]. 

\begin{figure} [ht!]
    \centering
   \includegraphics[width=01\linewidth]{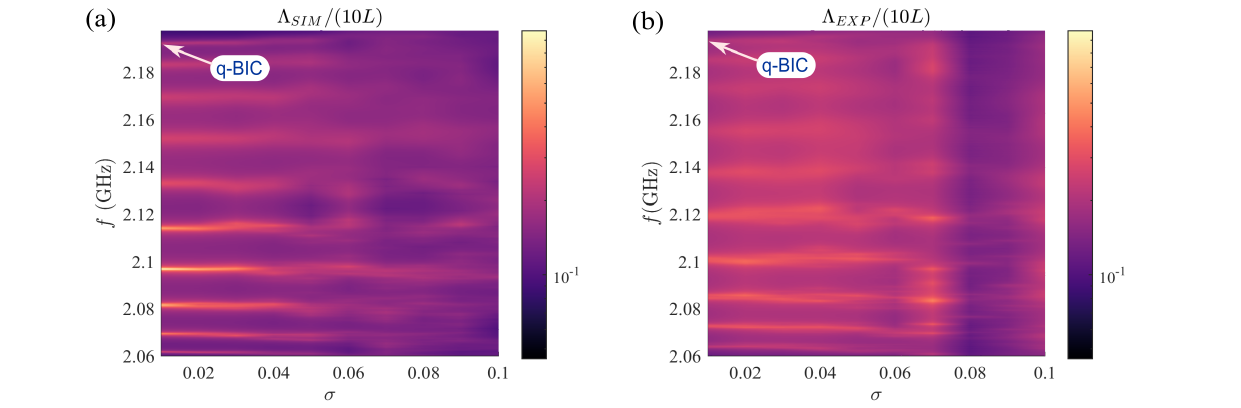}
     \caption{Numerically calculated (a) and experimentally obtained (b) dependencies of normalized localization length $\Lambda$ on the length of the chain $NL$  from the  disorder parameter $\sigma$ in a 10-disk system.}
    \label{ill}
\end{figure}

 We started from the calculation of transmittance spectrum in 10 chain of disks. The localisation length is obtained by the averaging of transmittance over the 10 realisations as it was done in experiment. The obtained results are shown at Fig.\ref{ill}. Since, if the wave is not localised, it completely passes from true the structure. Indeed, as it shown in Fig.\ref{ill} for the forbidden, e.i, waveguide modes frequencies one can see the maximums of the $\Lambda$, while for the frequencies above light line the values are small. Additionally, as the parameter $\sigma\geq 0.06$, the $\Lambda$ values become less distinct. This indicates the emergence of hybrid states, characterized by localized  features.

    {S1.} Maslova, E. E., Rybin, M. V., Bogdanov, A. A., \& Sadrieva, Z. F. (2021). Bound states in the continuum in periodic structures with structural disorder. \textit{Nanophotonics}, \textbf{10}(17), 4313–4321. De Gruyter.

    {S2.} Poddubny, A. N., Rybin, M. V., Limonov, M. F., \& Kivshar, Y. S. (2012). Fano interference governs wave transport in disordered systems. \textit{Nature Communications}, \textbf{3}(1), 914.